\documentclass[showpacs,amsmath,preprint,endfloats*,prb]{revtex4}
\usepackage{graphicx}
\usepackage{dcolumn}
\usepackage{bm}
\usepackage{amsmath,amssymb,amsfonts}
\usepackage{epsfig}

\begin{document}

\title{Theoretical estimation of density profile of semiflexible polymers adsorbed on a surface and thermodynamic glass transition scenario by means of the Bethe-Peierls approximation}
\author{F. Semeriyanov, A.I. Chervanyov, G. Heinrich}

\affiliation{Leibniz Institute of Polymer Research Dresden, Hohe
Strasse 6, D-01069 Dresden, Germany}

\date{\today}

\begin{abstract}
We develop a theory describing density profile of the
semi-flexible polymers absorbed onto a  planar surface. The
theoretical analysis consists of two parts. As a first part, we
calculate a density profile of the adsorbed polymers by developing an
extension of the Bethe-Peierls approximation to the case of
nonhomogeneous systems. This approach relies on  the combination of
the single chain adsorption theory and the lattice version of the
self-consistent field theory. Semi-flexibility of a chain is
described by incorporating a finite coordination number of the
lattice into the consideration, in the spirit of the previous
Silberberg approach. The developed lattice theory
incorporates the interaction between nearest-neighbor pairs of
segments and finite chain length. The theory is completely mapped into
the Scheutjens-Fleer theory in the limit of infinite coordination number. As a
second part of the developed approach, we calculate the configurational
entropy to investigate how the
density structure of the semi-flexible polymers near the surface
relates to possible reduction of glass transition temperature near
 nonabsorbing surface and enhancement near the strongly attractive surface.
\end{abstract}

\maketitle

\section{Introduction}
Despite extensive extensive experimental and theoretical work, some aspects of the polymer glass formation still remain unresolved puzzles. In the present work we address one such unresolved problem related to the role of the surface in the glass formation.
The glass transition in confined geometries has been in a focus of intensive experimental and theoretical investigations in several decades. Scientific disputes continue concerning whether depression \cite{DecreaseTg2, DecreaseTg3, DecreaseTg4, DecreaseTg5, DecreaseTg6, DecreaseTg7, SimulDecreaseTg, TheorDecreaseTg} or enhancement \cite{IncreaseTg1a, IncreaseTg, IncreaseTgSurfEng, IncrTgAttrSurf, InctrTgAtrrSurf, IncrTgNeutrSurf1, IncrTgNeutrSurf2, Saalwachter} of glass transition temperature $T_g$ is exerted by surface. The dynamic aspect of the slowing down of the cooperative molecular motions is attributed to the well-known mechanisms described by the classical theories. \cite{Adam_Gibbs, Cohen_Grest, Goetze} Despite some progress in understanding of the vitrification transition achieved with help of these theories, the driving mechanism of the cooperative motions near the glass transition is rather poorly understood. Restrictions imposed on these motions by the presence of interfaces in the system result in increasing of the enthalpy and reducing the entropy that can be described in terms of relevant characteristic length scales. According to the dynamic and calorimetric measurements, the glassy dynamics in sufficiently thin layers is rather significantly affected by the molecular weight of glass forming of polymers and layer thickness. \cite{Tress} The chain length is believed to affect $T_g$ only via the free volume associated with chain ends. \cite{Ferry, deGennes_glass} In this scenario the glass transition temperature is solely determined by length of typical loop adsorbed by both ends onto the surface. In the present study, the effect of chain length is investigated and the loops are taken into account. Note that in a typical experimental situation, the glass transition is considered to be primary caused by rapid cooling of the system below the glass transition temperature. We however consider the system at constant temperature fixed slightly below or above the bulk $T_g$ and study the thermodynamics of the system near the surface.

The above described need in the understanding of a physical nature of the glass transition in non-uniform polymer systems calls for an adequate theoretical approach to this important problem. The present work aims at providing such a theoretical model based on the detailed investigation of the density profiles of semiflexible polymers in presence of adsorption interactions with the substrate confining the polymer system. With the help of the developed approach, we intend to demonstrate that the non-uniform density structure of the aforementioned thin film underlines the specificities of the glass transition in these systems.

Despite the fact that glass formation is generally dynamical phenomena, the main features of the vitrification transition can be captured by applying the tools of equilibrium thermodynamics. An explicit recipe that allows one to relate the glass transition temperature with the configurational entropy $S_{conf}$ of glass forming polymers has been proposed by Gibbs and DiMarzio \cite{Gibbs-DiMarzio}. In order to make quantitative connections between the glass transition and polymer density gradients in thin films, we intend to explore "Kauzmann paradox" or entropy crisis to occur in the nonuniform systems. Recall \cite{Kauzmann1948} that this paradox predicts a possible configurational entropy vanishing near the glass transition point that has been resolved in Gibbs-DiMarzio theory by the following scenario: Due to rapid precipitation of the configurational entropy upon cooling at $T_K \approx 0.77 T_g$ and crossing through the dynamical transition point, no thermodynamically stable state can exist below this point so that the system falls out of equilibrium. In non-uniform systems, similar thermodynamically unstable regions may appear in the domains of enhanced density thus causing vitrification in these domains. Upon rapid cooling in presence of density gradients caused by presence of confined surfaces, a non-uniform polymer system can therefore be subdivided into the glassy and melt domains. This effect can explain the experimentally observed difference in the role of confining surfaces in the glass formation.

In the present study we address above described effects by developing a lattice model
that takes into account the semiflexibility of a chain as well as the interactions with other chains near the adsorbing surface. One of the main advantages of the developed approach is that it goes beyond the Bragg-Williams athermal limit incorporated into the Flory-Huggins model underlying most of the current theories \cite{DoiEdwards, dG, Roe, Scheutjens-Fleer} of the non-uniform polymer systems. In particular, the mentioned previous theories do not take into account the finite semi-flexibility of the chain that is one of the main issues in the present study of the polymer structure near the
adsorbing surface. The developed lattice model allows for a more
straightforward incorporation of the semi-flexibility of chains in the many chain problem with excluded volume taken into the consideration, in comparison with the continuous theories based
on the worm-like chain model.\cite{Daoulas, Schmid, Ganesan} In
addition, our model provides more realistic description of the chain
adsorption problem due to the fact that it operates with the actual
binding point interaction between the polymer segments and the
adsorbing surface, rather than the extended adsorption potential
that adequately describes only the case of physical polymer
adsorption. We believe that the above features make the present
approach valuable contribution to the field of studying the density
structure of non-uniform polymer systems.

Despite a long history of using lattice-based models for the
description of the polymer adsorption onto a planar surfaces, only
few of this models can be credited for giving an adequate
description of the polymer segment-segment interactions. Most
advanced of these models are based on the approaches of
Hoove,\cite{Hoove} Silberberg,\cite{Silberberg} Roe,\cite{Roe} and
Scheutjens and Fleer (SF).\cite{Scheutjens-Fleer} Each of these
models have severe limitations, SF being the most advanced and
adequate for describing many relevant experimental situations. The
main advantage of the SF model over the approaches of Roe and
Silberberg is in the provided possibility to describe the actual
distribution of \emph{finite} chains on a lattice in the presence of
an external field. This theory, however, relies on the Flory-Huggins
model \cite{Flory, Huggins} for the enthalpy that is known \cite{GujratiJCP, dG} to be
non-adequate for the description of many relevant phenomena in
non-uniform polymer melts. In addition, to the best of the authors
knowledge, SF theory in its conceptional form has never been used to
describe the finite stiffness of the chains.

Along with using more advanced in terms of correlations taken into account approximation, the present study remedies this omission by using the modified lattice coordination number that
distinguishes between trans and gauche configurations of the polymer
chain, as has been suggested previously by
Silberberg. \cite{Silberberg} In our former study,\cite{2011} we
applied the Bethe-Peierls approximation \cite{Bethe, Peierls} to the study of the uniform
polymer melts and demonstrated the complete agreement of this
approach with  quasi-chemical approximation. The present work
describes an extension of our earlier studies to the case of
non-uniform polymer melts  in the presence of adsorbing surface.
Specifically, we apply the developed formalism for deriving the
adsorption isotherm in the described
non-uniform many-chain system, which allows us to mimic the
density structure of the whole polymer film. Upon analyzing the density
structure of the adsorbed semi-flexible chains, we perform a
numerical estimation of the configurational entropy of the
non-uniform chain layer. In performing these calculations, we assume
that polymers experience ideal glass transition at the point where
the configurational entropy vanishes.  The described
model mimics the the apparent glass transition in dense systems.
As we intend to show in what follows, this relatively simple model
provides an effective tool for studying the glass transition in
confined geometries.

The paper is organized as follows. In  Section  \ref{SecTheory} we describe the
mathematical procedure that rationalizes the mentioned
generalization of the Bethe-Peierls approximation to the case of
nonhomogeneous chain systems. In Section \ref{GlasTransSec} we introduce the
calculation method of the configuration entropy along with the
equations for the contact densities. In Section \ref{SecResults} we present the calculated density
profiles, along with the corresponding values of the configurational entropy given as a function of the  bulk polymer density observed
far away from the adsorbing surface.  Finally, Sec. \ref{SecConcl} presents our conclusions.

\section{Density profile of semi-flexible chain in presence of
adsorbing surface: Mathematical model} \label{SecTheory}

In performing the first step of the program outlined in the
Introduction, in this Section we intend to calculate the density
profile of semi-flexible chains adsorbed onto a planar substrate. For the sake of notational convenience we distinguish between the segments positioned at the ends of a chain and the rest of the segments that we call hereafter the end segments and the middle segments, respectively. The adsorption interaction between polymer segments and the substrate is described by the binding energy $U_s<0$ that can
be alternatively characterized by the characteristic adsorption
length $L_A\sim (\beta U_s)^{-1}$, $\beta$ being the reciprocal
temperature expressed in units of the Boltzmann constant $k_B$. For
the sake of simplicity, we assume that the total width $D$ of the
adsorbed film is much larger than $L_A$, so that the profile of the
chain density in the described film comprised of the chains
saturates to its bulk value observed far away from the adsorbing
surface.

Note that there are essential differences between the behavior of the end segments and middle segments caused by the fact that confining surface restricts the location of these types of segments in different ways. First restriction on the possible locations of the both types of segments is caused by the limitations imposed by the surface on the coordination number. The restriction equally applies to both types of segments. The second restriction caused by the connectivity of chains, in contrary, differs for the end and middle segments. Specifically, the configuration with the end points placed near the surface makes this configuration more entropically favorable than the equivalent configurations with middle segments placed in the same positions. the above difference can be quantitatively taken into account by straightforward modification of our previous theory \cite{2011} developed for uniform polymer systems. Technically, this modification amounts to introducing the correction to the configurational entropy differentiating the end and middle points of the polymer chains described in what follows.

Owing to the added mathematical complexity in describing
semi-flexible polymer systems, far less is known about the density
profiles of the adsorbed chains that possess a finite stiffness or
rigidity. In the present study, we will take the stiffness of a
chain into account by employing the Flory lattice model
Ref. \onlinecite{2011}. The essence of the approximation is to distinguish between trans and gauche states of the model chain: The
latter can pass from any lattice site to the directions
of the nearest-neighbor sites. The forward direction, the trans
state, is energetically more favorable to all other directions, the
gauche states. The energy difference $U_f$ between the trans
and the gauche state, introduces a local stiffness so that at
temperature $T = 0$, the chain will be in the all trans state. At
any finite temperature, because of thermal perturbations, a
long polymer will be in a coiled state, same as for a completely
flexible polymer when $U_f = 0$. A first order phase transition
was found to take place \cite{Flory1956} in the many chain system at a constant
density when the excluded volume interactions are taken into
account. In the framework of this theory, the
persistence of a chain can be described by the flexibility energy
$U_f$, whereas the finite thickness of this chain is associated with
the size of the lattice cell. Note that the proposed lattice model is restricted to only nearest-neighbor interactions among segments described by the potential $V$ that has the same strength irrespective whether nearest neighbors are connected by a bond or not.
Our polymer melt possesses a finite compressibility captured by means of voids distributed in such a way that a constant
chemical potential is maintained throughout the system. Apart from the first layer, $z$ is assumed
to be independent of distance from the adsorbing surface. In addition, the stiffness of the model chains is considered to be independent of the local chain
density in the system.

\begin{figure}
\epsfxsize=3.2in \epsffile{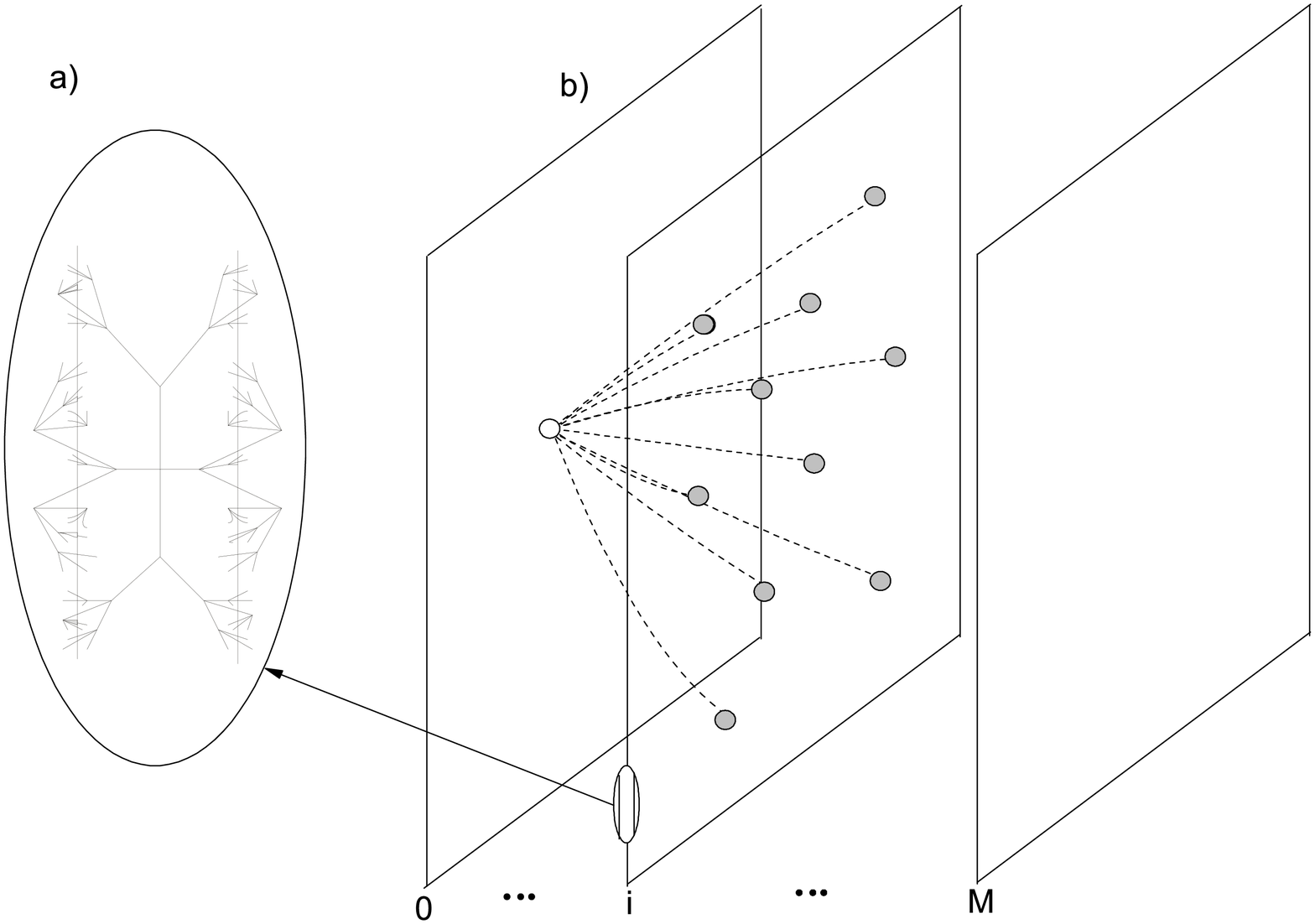}%
\caption{a) Cayley tree of 4 generations with coordination number $z=4$, constricted to within 1+2 consecutive layers. b) Schematic picture of the layer structure and connections of Cayley tree sites with a ghost site.} \label{Fig1}
\end{figure}

A sketch of the used Cayley tree terminated after the 4-th generation step is shown in Fig. \ref{Fig1}a. This Cayley tree is embedded into the lattice that consists of a sea of Cayley trees connected to each other through their termination sites. The present theory, initially developed for homogeneous polymers, relies on
establishing a connection between the local configuration of species occupying lattice sites of the Cayley tree and an average local field entering into the grand partition function (GPF) of the system. This relation is derived by choosing the local configuration
weights according to a certain field relation between internal and
external interactions in the neighborhood of the central site of the Cayley tree. For the considered case of the layered structure in
the presence of adsorbing surface, this relation and resulting GPF
must be separately determined for each layer of the system. In order to construct a layered structure in the proposed formalism, we embed the Cayley tree into 3 consecutive layers as shown in Fig. 1b. Finally, the equal probability condition \cite{Roberts, Chang1939} of the site occupation on the Cayley tree is applied to that particular structure. Using the recursive relations between consecutive layers makes it possible to estimate the
density profile in the film of semi-flexible chains adsorbed onto the surface.

Let $L$ denote the number of sites in a layer, $\theta_{i}L$ the number of occupied lattice sites, and write the Boltzmann weight for
the interactions between occupied sites $\eta = \exp( -V / k_B T)$, the absolute activity $\lambda = \exp (\mu/k_B T)$, $\mu$ being chemical potential; $w_{f}=\exp (-U_{f}
/k_{\text{B}}T)$ and $w_{s} = \exp (-U_s/k_B T)$ the Boltzmann
weights for the flex energy and adsorption energy, respectively. The configurational energy of the system is $E_{i}=(VN_{11} + U_{f}N_{g} + U_{s}N_{s})_i$, where $N_{11}$ is the number of nearest neighbor pairs of segments irrespective to wether they form bond or not, $N_{g}$ the total number
of the gauche configurations, $\mu$ the chemical potential. The grand partition function of the system can be written as
\begin{equation}\label{GPF_total}
    \Xi = \sum_{\{\theta_{i}\}} Q(\{\theta_{i}\}) \lambda^{\theta_{i}L} =  \sum_{ \{\theta_{i}\}} \sum_{E} \Omega(\{\theta_{i}\},E) \lambda^{\theta_{i}L} \exp (-E/k_{B} T) ,
\end{equation}
where $Q$ is the canonical partition function. The series can be replaced by its maximum term with only negligible error. It means that among all sets $\{\theta_{i}\}$'s one chooses only the set $\{\theta_{i}^{\ast}\}$ that corresponds to the equilibrium density in the given layer, $\theta_{i}^{\ast}$.
The equilibrium grad partition function of a single layer $Z^{\ast}_{i}$ that corresponds to the equilibrium $\theta_i^{\ast}$ in the layer $i$ is given by
\begin{equation}
Z_i^{\ast}=\sum_{\theta_{11},\,\theta_{g},\, \theta_{s}} \Omega (\theta_{i}^{\ast},\theta_{11},\theta_{g},\theta_{s}) \eta^{\theta_{c,i}L} w_{f}^{\theta_{g} L} w_{s} ^{\theta_{s} L},
\label{GPF}
\end{equation}
where $\Omega $ is the number of arrangements on a layer of $L$ sites
that have the same densities $\theta_{11}(\theta_i^{\ast})$, $\theta_{g}(\theta_i^{\ast})$, $\theta_{s}(\theta_i^{\ast})$.

The standard method to calculate the equilibrium grand partition function is to use the Bethe-Peierls approximation. \cite{Chang1939} The starting point is to take the Cayley tree embedded between 3 consecutive layers with the central site site placed in the layer $i$ surrounded by $z$ first generation
sites situated in the same layer. This structure is thought to represent an aggregate of segments in the sea of molecules constituting the polymer melt. For the sake of tractability, we assume that the Cayley tree structure is maintained outside the aggregate meaning that this structure possesses no
closed loops at least for the $x$ number of generations sufficient to embed a single chain. In order to denote the state of
occupation of the above $z+1$ sites, we define a set of numbers
$\{\delta _{j}\}=\delta _{0},\delta _{1},...,\delta _{z}$ that
assume the values of  0 or 1 for empty and occupied sites,
respectively. The index $j=0$ denotes the central site, while the
indexes $1\leq j\leq z$ denote the nearest-neighbor sites. The
location of the aggregate and its contribution to the total
partition function is therefore uniquely defined by the position of
its central site. In particular, the configuration where the central
site of a specific aggregate is unoccupied $(0;\delta
_{1},...,\delta _{z})$ contributes to GPF as the sum over all the
arrangements of the nearest neighbors
\begin{equation}
\label{sum}
   Z_i^{0}= \underset {\{\delta_{j}\} {\sum}} \lambda ^{\delta _{1}+\cdots +
   \delta _{z}}\psi (1,...,z),
\end{equation}
where $\psi$ is yet unknown function that corrects for various
configurations of the molecules outside the aggregate. Since $\psi$
cannot be determined analytically, one has to resort to simplifying
approximations of this function due to Bethe and Guggenheim. \cite{Miller}  According to the mentioned approximation, $\psi$ represents the mean-filed superimposed on
the first generation sites of the Cayley tree and is given by
\begin{equation}
\label{psi}
     \psi =c\xi ^{\delta _{1}+\cdots +\delta _{z}},
\end{equation}
where $c$ is  constant, $\xi$ is the unknown field that is to be
related to the occupation probabilities on the lattice corresponding
to the specific layer. The contribution to GPF from the
configuration $(0;\delta_{1},...,\delta _{z})$ follows immediately
from Eqs.(\ref{sum},\ref{psi}), to be written as
\begin{equation} \label{GPF_sm_cluster}
    Z_i^{0}=\underset{\{\delta_{i}\}} {\sum } \psi = c
    \Psi_i^z,
\end{equation}
where $\Psi_i=1+\lambda\xi_i$.  As we have shown in our previous
work, \cite{2011} the derived approximation for the partial partition
function is equally applicable to the Cayley tree of $x$ generations capturing a chain connected to its $z^{\prime}=2(z-1)+(x-2)(z-2)$ nearest neighbor sites.

Determining  the contribution to GPT of the Cayley tree with the
occupied central site is considerably more mathematically demanding problem. The key element of the problem is to estimate the conditional probabilities of the occupancy
of the nearest neighbor sites by other segments of the same chain
molecule. In order to rationalize the above correlations among the
occupancies of the aggregate sites imposed by the connectivity of
the chain, we define the partial partition functions (PPF) $\chi _{k}(h)$. This PPF describes configurations of a part of the chain consisting of $k$
bonds extending outside a first generation sites of the Cayley tree and $i$ being the number
of layers that separate the selected layer from the adsorbing
surface. In terms of functions $\chi _{k,i}$, the contribution to the
partition function coming from the aggregates with the occupied
central site can be written \cite{2011} as
\begin{equation} \label{small_agg_GPF}
Z_i^{1}= c z \Psi _{x,i} ^{z-1} \lambda^{2} \eta \chi _{x-1,i} + 2^{-1}
r_{f} c z
 \Psi _{x,i} ^{z-2} \lambda^3 \eta^2 \underset{k=1}{
\overset{x-2}{\sum }}\chi _{k,i}\chi _{x-k-1,i},
\end{equation}
where $\Psi_{x,i} \equiv (1+\lambda \eta \xi_{i} )$. Here, the first term in the first term in the right hand side
(r.h.s.) corresponds to the endpoint residing in the central site taking into account all $z$ possibilities for the first bond to extend from from the center of the Cayley tree. The second term takes into account the case when the central site is occupied by the middle segment with the coefficient
$r_{f} = w_{f}(z-2)+1$ accounting for the finite stiffness of the
chain, i.e. the latter weights $z-2$ out of $z-1$ possible configurations with the the Boltzmann weight $w_f$. The total GPF of a single layer separated from the adsorbing
surface by  $i$ lattice steps  is given by a sum of the partition
functions given by Eqs.(\ref{small_agg_GPF}) and
(\ref{GPF_sm_cluster}). Note that in contrast to
Eq.(\ref{GPF_sm_cluster}), Eq.(\ref{small_agg_GPF}) incorporates the
parameter $\eta$ that describes the nearest-neighbor interactions
between the segment placed into the central cite and the other
segments that belong to the same aggregate. Note that  the index $i$ is not included in the presentation for further convenience but assumed implicitly in what follows for all layer dependent quantities.

In order to rationalize the above derived contributions to the
partition functions one must rely on some assumptions about
the structure of the lattice surrounding the aggregate, as described by the fields $\xi$ and $\chi_k$. The model is appropriate for describing the considered system by the Cayley tree with the number of generations of brunches not larger than the number of polymer bonds, which takes into account local configurations. Non-local configurations are taken into account
using the axillary fields $\xi$ and $\chi_k$ resolved by the equal
probability condition \cite{Roberts, Chang1939} that the occupancy state of the central site is equal to that of sites of the first generation sites of the Cayley tree. We will elaborate on that in what follows. The introduced partial partition function $\chi_k$ that describes
the correction to the partition function beyond its dimeric part for the outside part of chain has been previously
calculated for the homogeneous system \cite{2011} using the above tree structure in analogy with the homogeneous Bethe lattice theory \cite{Ryu-Gujrati1997}:
\begin{equation}\label{hom_chi}
    \chi_k = p\chi_{k-1},k\geq1,
\end{equation}
where $p$ is the probability to have one extra bond extending beyond the center of the Cayley tree, $p = r_f \lambda \eta \Psi_x^{-1}\chi_1$ and $\chi_1 = \Psi_x^{z-1}\Psi^{-(z-1)}$; with the power $(z-2)$ applied to $\Psi_x$ due to having $z-2$ local gauche conformations out of $z-1$ possible for a middle segment located anywhere except the center of
the Cayley tree, and the power $-(z-1)$ is applied to $\Psi$ that is being deduced \cite{2011} from the dimer theory\cite{Chang1939} by taking the limit $x=2$.
For the purposes of the present work, this relation is to be extended to the case of non-uniform systems. We observe that the definition
\begin{equation}\label{chi_0}
    \chi_0 \equiv p^{-1} \chi_{1}
\end{equation}
allows to incorporate both terms in (\ref{small_agg_GPF}) to the same sum with extended limits. Formally, $\chi_0$ corrects for the absence of the bond in the end point of the polymer chain.

In order to incorporate the above mentioned difference between the behavior of the end and middle segments near the adsorbing surface into the above lattice model one has to resort to the standard method of percolation theory \cite{Stinchcombe, Reynolds} on the Cayley tree that describes critical phenomena by making use the concept of "ghost site" (GS). This construction is illustrated in Fig. \ref{Fig1}b. As can be seen GS is simultaneously connected to all sites of the tree with probability $h$ per site. This concept can be rationalized by introducing surface auxiliary field $h$ that describes the corrections to the probabilities of the configurations due to the presence of the surface. The presence of GS modifies the form of the bulk PPFs that can be written as
\begin{equation}\label{inhom_chi}
    \chi_k(p,h) = \chi_k(p)(1-h)^{k+1},
\end{equation}
The case of the bulk PPF can be retrieved from (\ref{inhom_chi}) taking $h=0$. Non-zero $h$ corresponds to the situations where each site of the chain is connected to the GS which mimics polymer-surface interaction.

In order to determine $h$, it is necessary to consider the quantity $\chi_x(p,h)$ representing the probability to have simultaneously $x$ sites and bonds comprising the chain. The configuration described by $\chi_x(p,h)$ violates the above property of the Conventional Cayley tree of not having closed loops. The formation of such loop is solely due to the presence of the GS. For this special configuration at least one site of the Cayley tree occupied by the chain has to be connected to GS, which corresponds to $h=0$. Owing to the absence of the end-points in the closed loops $\chi_x(p,h)$ factorizes into $x$ probabilities $p$ of the bond formation. Thus simple relation  $\chi_x(p,h)=p^{x}$ allows us to define $h$ as an analytical continuation of $\chi_k(p,h)$ given by (\ref{inhom_chi}) by extending the recursive relation (\ref{hom_chi}) to the case $k = x$ to yield
\begin{equation} \label{field_h}
    1-h = (p^{2} \chi_{1}^{-1})^{1/(x+1)}.
\end{equation}
In addition, we define $P_k(p,h) \equiv \chi_k(p,h)/\gamma^{1/2}$, where $\gamma$ is the normalization factor that is introduced to have the bulk properties far away from the surface, which is represented by
\begin{equation}\label{gamma_norm}
    \left(\sum_{k=0}^{x-1}P _{k} P _{x-k-1}\right)_b=x,
\end{equation}
Using the latter definition in (\ref{small_agg_GPF}), one finds that
\begin{equation} \label{small_agg_GPF1}
Z = c\Psi^{z} + c\frac{z}{2} r_{f}^{-1}  \lambda \Psi _{x} ^{z} p^{-2} \gamma \underset{k=0}{ \overset{x-1}{\sum }}P _{k} P _{x-k-1},
\end{equation}

The above arguments about the connection between homogeneous and non-homogeneous partial partition functions can be generalized to to the recursive relations on the Cayley tree occupying 3 consecutive layers with the same ghost field $h$ depending on number of steps from the surface:
\begin{eqnarray} \label{req_rl_P}
P _{0} &=& (1-h)\chi_0/\gamma^{1/2}, \nonumber \\
P _{1} &=& (1-h)p (\lambda_0 P_0 + \lambda_1 P_{0}^{(-)} + \lambda_1 P_0^{(+)}), \nonumber \\
&&\vdots \\
P _{k} &=& (1-h)p (\lambda_0 P_{k-1} + \lambda_1 P_{k-1}^{(-)} + \lambda_1 P_{k}^{(+)}),  \nonumber
\end{eqnarray}
Here, the coupling constants $\lambda_1$ and $\lambda_0$ weight contributions from the three neighboring layers and $\lambda_0 = (1-\lambda_1)/2$ is the fraction of the average coordination number $r_f$ to come back to the same level at each step. The superscripts "$-$" and "+" denote the lower and upper layers, respectively. These recursive relations describe the fact that the chains occupying these 3 coupled layers are connected to the ghost site with the probability $h$, see Fig. \ref{Fig1}. More precisely, a $(k-1)$-part of the chain residing have a chance to continue on the next step to one of 3 layers with the probability determined by $p$ and the corresponding coupling constant. The equations (\ref{req_rl_P}) serves as the implicit set of equations for determination of the layer dependent occupation, $\theta_i$, by means of the numerical iteration. 

The layer dependent density $\theta$ is obtained as the ratio of the term in the GPF
corresponding to the occupied central site of the Cayley tree to the total GPF, $\theta = (Z-c\Psi^{z})/Z$, which gives, upon substituting into (\ref{small_agg_GPF1}), an important relation:
\begin{equation} \label{x-mer eqs bA}
 \frac{z}{2} r_{f}^{-1}  \lambda p^{-2} \gamma
 \underset{k=0}{ \overset{x-1}{\sum }}P _{k}P _{x-k-1}   = \frac{\theta }{1-\theta } \frac{\Psi ^{z}}{\Psi _x ^{z}}.
\end{equation}
In order to calculate the auxiliary field $\xi$ we employ the equal probability condition to find a void in the center of the Cayley tree or in the arbitrary site belonging to its first generation. In terms of $\xi_k$ this condition reads
\begin{eqnarray} \label{cond2}
Z^{-1}c\Psi ^{z} &=& Z^{-1}\{c\Psi^{z-1} + c (z-1) \Psi _{x} ^{z-2} \lambda
^{2} \eta \chi _{x-1}  \nonumber \\
&&+ c\frac{z-2}{2} r_{f} \Psi _{x} ^{z-3}\lambda ^{3}\eta
^{2}\underset{k=1}{ \overset{x-2}{\sum }}\chi _{k}\chi _{x-k-1}\},
\end{eqnarray}
where the quantity in the curly brackets is the partial partition
function for the first generation sites calculated under the condition
that one such site is empty.

Using (\ref{x-mer eqs bA}), we rewrite (\ref{cond2}) in a more tractable form
\begin{equation} \label{x-mer eqs bB}
    \varepsilon = \left( X\delta -1 + \sqrt{(X\delta-1)^2+4\eta X\delta} \right)/2\eta.
\end{equation}
where $\varepsilon \equiv \lambda \xi$ and $X \equiv \theta/(1-\theta)$ and $\delta$ is given by
\begin{equation} \label{delta_eq}
    \delta =  \frac{z-1}{z} -
\frac{1}{z} \underset{k=1}{ \overset{x-2}{\sum }} P_k P_{x-k-1} \left/ \underset{k=0}{ \overset{x-1}{\sum }} P_k P_{x-k-1} \right.
\end{equation}

In the limit of large distances from the surface the bulk properties should be attained. The latter provides the value of the bulk segment activity $\lambda$ obtained in Ref. \cite{2011} and reproduced here for convenience:
\begin{eqnarray}
    \lambda &=& p_b(r_f \eta)^{-1} (\Psi^{z-1}/\Psi_x^{z-2})_b, \label{lambda_eq_st} \\
    p_b &=& [2^{-1} z r_f^{-2} \eta^{-1} x X_b^{-1} (\Psi_x^2/\Psi)_b]^{-1/x}
\end{eqnarray}
where the subscript "b" stands for the bulk and the $\Psi$-factors are determined from (\ref{x-mer eqs bB}) with $\delta_b = z^{\prime}/zx$. Substituting the above bulk chemical potential into (\ref{x-mer eqs bA}) and making use of (\ref{gamma_norm}), the normalization factor $\gamma = p_b^{x+1}$ is found and finally this leads to
\begin{eqnarray}
 X &=& 2^{-1}zr_{f}^{-2} p_b^{x+1} \eta^{-1} \Psi_x^2 \Psi^{-1} p^{-1}
 \underset{k=0}{ \overset{x-1}{\sum }}P _{k}P _{x-k-1}, \label{theta} \\
 p &=& p_b w_s (\Psi_x^{z-2}/\Psi^{z-1})(\Psi^{z-1}/\Psi_x^{z-2})_b, \label{p}
\end{eqnarray}
where $w_s$ differs from unity only for the first layer near the surface. Additionally, a special care must be exercised  regarding in the first layer where the reduced coordination number, $(\lambda_0+\lambda_1)z$, should be enforced due to absence of the lower layer next to the surface. This concerns quantities $p$ and $\delta$ in the equations above. The recursive relations (\ref{req_rl_P}) together with eqs. (\ref{lambda_eq_st}) and (\ref{theta}) constitute the system that describes the density profile near the surface. 

The so-called surface excess $\Gamma$, the average number of polymer segments per unit area of the surface in excess of those present in the bulk, is defined by
\begin{equation}\label{alpha}
     \Gamma = \sum_{i=0}^{M} (\theta-\theta_b).
\end{equation}
This quantity is useful to estimate the surface tension of a polymer solution by taking the random mixing approach. \cite{Scheutjens-Fleer}

The Bragg-Williams approximation is attained by means of the proper limit, $z \rightarrow \infty$, $Vz=const$, that produces the following expression for the chemical potential $\mu = \frac{1}{x}\ln \lambda$ of the semiflexible polymer melt:
\begin{equation} \label{mu_FH}
    \mu_{BW} = \frac{ 1}{x} \ln \frac{\theta_b}{x} +\chi_f \ln r_f - \ln (1 - \theta_b) -  2\chi  \theta_b.
\end{equation}
where $\chi = z (\eta - 1)/2$ is the Flory-Huggins chi-parameter, $\chi_f \equiv -x^{-1}\ln(0.5zr_f^{-2})$ is a chain stiffness parameter.

The Scheutjens-Fleer (SF) theory \cite{Scheutjens-Fleer} is recovered by employing the Brag-Williams approximation and using the condition of layer independent chemical potential. When these conditions are applied to (\ref{p}) supplemented by (\ref{x-mer eqs bB}), one finds that
\begin{eqnarray}\label{Delta_U}
 \ln p &=& \frac{1}{x} \ln \frac{\theta_b}{x} +\chi_f -\Delta U, \\
 \Delta U &=& -\ln \left( \frac{1-\theta}{1-\theta_b} \right) - 2\chi(\theta-\theta_b) - (\chi_s+\lambda_1 \chi) \delta_{i,1},
\end{eqnarray}
where $\chi_s = -U_s/k_B T$ is the adsorption energy with $\delta_{i,1}$ being the Kronecker delta selecting the first layer near the surface. Performing further simplifications of (\ref{theta}) with the help of (\ref{x-mer eqs bB}), we obtain an expression identical to the corresponding expression in the SF theory:
\begin{equation} \label{theta_SF}
 \theta = \theta_b e^{\Delta U} x^{-1} \underset{k=0}{ \overset{x-1}{\sum }}P _{k}P _{x-k-1}.
\end{equation}

\section{Glass transition} \label{GlasTransSec}
Theoretically, the problem can be approached from the prospective of thermodynamic theory of glass transition by Gibbs and DiMarzio \cite{Gibbs-DiMarzio} and the dynamic extension of Adam and Gibbs \cite{Adam_Gibbs}. The thermodynamic origin of the glass transition was motivated by Kauzmann's discovery that the entropy of equilibrium liquid supercooled below the glass transition temperature becomes less than the entropy of crystal at the same temperature. \cite{Kauzmann1948} Discussions about Kauzmann paradox continue since its discovery till now. In particular, it is attributed to the existence of a thermodynamic origin in the vitrification phenomenon based on the concept of ideal glass. According to the scenario, a substance passing from liquid to glass experiences the influence of the ideal glass state in terms of the rapid reduction in number of configurational states available to probe. Thus, the glass transition is associated with a confinement in the configurational space caused by proximity to the ideal glass state. Owing to extremely long relaxation times of glassy motions, this model is difficult to verify experimentally or computationally. However, indirectly, such scenario finds its support in experimental fact that dependence of relaxation time $\tau$ on temperature is well-fitted by the Williams-Landell-Ferry (WLF) law $\ln \tau(T)/\tau(T_g) \sim (T-T_0)^{-1}$, where $T_0 \sim T_K$ corresponds to the temperature at which infinite relaxation time is expected. \cite{Dalal1, Dalal2} The existence of such point with infinite viscosity speaks in favor existence of a unique configuration that cannot be reached by any experimental technique.

In order to find the entropy expression beyond the random mixing, we obtain the layer configurational entropy by means of the integration of the absolute activity $\eta$, which is the technique first employed by Guggenheim \cite{Miller} for the homogeneous problem and extended by us to the nonhomogeneous system as follows: 
\begin{equation} \label{Sconf}
S_{conf}^{(i)}/k_{\text{B}} = -\underset{0}{ \overset{\theta_b}{\int }} \ln [\lambda(\theta_b)] \theta_{i}^{\prime}(\theta_b) \text{d}\theta_b -\theta _{11}\ln \eta -\theta _{g}\ln w_{f} - \theta_s \ln w_{s},
\end{equation}
based on the known dependence of $\lambda$ on $\theta_b$ given by (\ref{lambda_eq_st}).
The surface contact density $\theta_s$ is is equal to the density of the first layer, whereas the contact densities $\theta _{11}$, $\theta_g$, calculated in our previous work \cite{2011} to be
\begin{eqnarray}
\theta _{11}&=& z\theta_{i} -\frac{z}{2}+\frac{z}{2}\frac{(1-\theta_{i} )}{1+\varepsilon_i
},  \label{cont dens} \\
\theta _{g}&=&\frac{f \theta_{i} (x-2)}{x} .
\label{gauche dens}
\end{eqnarray}
can be shown to apply to the nonhomogeneous system as well.

In order to calculate the configurational entropy as a function of density, eqs. (\ref{req_rl_P}) and (\ref{theta}) are to be solved numerically for $\theta_i$ as a function of $\theta_b$ in each layer. The expression for the chemical potential (\ref{lambda_eq_st}) that relates the layer dependent $\theta_i$ and the parameter $\varepsilon$ through (\ref{x-mer eqs bB}) is also included into consideration. The slope of the function $\theta_{i}(\theta_b)$ is obtained in each layer to perform the numerical integration according to (\ref{Sconf}) to obtain the entropy in each layer.

\section{Results and discussion} \label{SecResults}
We obtained the density profiles by means of the iterative technique using the Roe \cite{Roe} theory as an initial guess of our iterations with $P_0=0$ and $P_M=1$ being the surface and the bulk values, respectively. The values of $\lambda_0$ and $\lambda_1$ were chosen to be $0.5$ and $0.25$, respectively, the same as in Ref. \cite{Scheutjens-Fleer}. The iterations does not require any additional features for the case of small coordination numbers and short chain lengths. The density is obtained from the the equation of state (\ref{lambda_eq_st}) for the case of finite coordination number $z=3$, and the fixed chemical potential and absolute activity, $\mu=-0.7$ and $\eta=1.04545$, respectively.

\begin{figure}
\epsfxsize=3.2in \epsffile{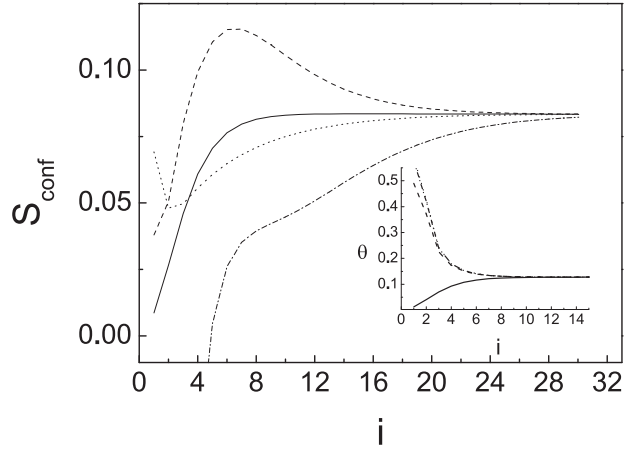}%
\caption{Profiles of configurational entropy of various adsorption energy: solid line - $\chi_s = 0$, dash line - $\chi_s = 0.5$, dotted line - $\chi_s = 0.56$, dash-dot line - $\chi_s = 0.57$. Other parameters of the model: $x=1000$, $f=0.5$, $z=3$ , same for all curves. The inset shows density profiles corresponding to the curve of the main figure.} \label{Fig_Conf_entr1}
\end{figure}

Profiles of configurational entropy of completely flexile chains are presented in Fig. \ref{Fig_Conf_entr1}. Specifically, the model contains a special feature that the configurational entropy becomes negative near attractive surface. This problem is known as Kauzmann paradox associated with the experimental observation of such entropy paradox below to the glass transition point. In the present work, we assign the ideal glass state with the point where the calculated configurational entropy becomes zero and consider the states with negative configurational entropies to be unphysical solutions of the model. The latter states are obtained only in the case when the attraction between surface and polymer segments is strong enough to achieve sufficiently high density near the surface, see inset of Fig. \ref{Fig_Conf_entr1}. On the other hand, the repulsive surface plays a suppressive role for the vitrification since entropies near the surface are strictly positive in that case, even when the bulk polymer happens to be below its bulk $T_K$ (not shown). Additionally, we observe that the configurational entropy being positive is always depressed near the repulsive surface, which is expected since surface reduces number of possible conformations of the chain. An attractive surface may induce both reduction and enhancement of $S_{conf}$ depending on the strength of attraction. Remember that the glassy dynamics is a property of entire film rather than the local dynamics at some distance away from the surface as discussed in Ref. \cite{GlassTransCollEff}. Thus, the proximity of the $S_{conf}$ to zero in the first layers, which is supposed to slow down the dynamics according to Adam-Gibbs theory, may also lead to the slowing down in the subsequent layers. In general, we find that profiles of configurational entropy are very sensitive to the surface potential and that they may alter dramatically by a slight change of the surface potential. As a example, we present Fig. \ref{Fig_Conf_entr1} demonstrating the above mentioned sensitivity to the change of $\chi_s$ from 0.56 to 0.57 leading to the appearance of negative entropies. A large discrepancy in experimental results regarding glass transition near surface mentioned in Introduction could originate from high sensitivity of configurational entropy to the system parameters. We also find that the Kauzmann paradox is generally related to high contrast between surface and bulk density achieved by means of applying high surface potential and low bulk density. In the course of the numerical calculation, one observes that there is an upper limit of surface potentials, $\chi_{s,max}\simeq 0.6$, above which the program does not converge to any physical solution. The nature of the latter is in presence of discontinuity in the surface coverage $\theta_1$ as a function of surface potential $\chi_s$. This effect originates from the presence of the attractive forces between polymer segments that accelerate the adsorption with increase of the surface coverage. The problem can be avoided by taking another boundary condition at the surface, which we plan to investigate in forthcoming publications. It is interesting that the Kauzmann paradox appears as a precursor to that transition at lower $\chi_{s}$ for the sufficiently long chains, $x=84$ ($\mu=-0.7$, $\eta=1.04545$, $f=0.5$, $z=3$), which can be attributed to the formation of a glassy layer near the surface in accordance with the experimental observations. For short chains the critical adsorption occurs with the sharp increase of the convergence time, but this point is not prelimited by the Kauzmann paradox in configurational entropy.

\begin{figure}
\epsfxsize=3.2in \epsffile{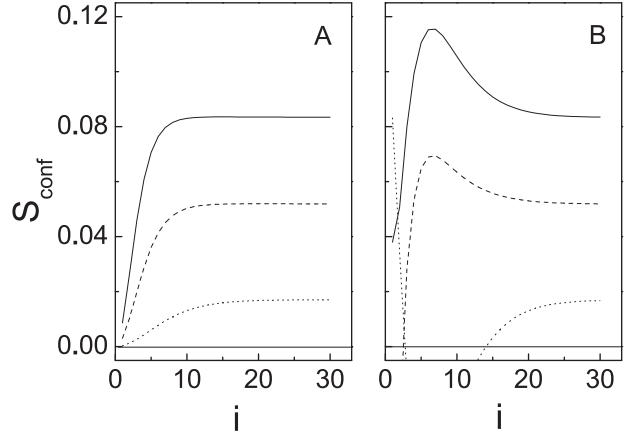}%
\caption{Profiles of configurational entropy for 3 values of gauche state probability: solid line - $f=0.5$, dash line - $f=0.49$, dotted line - $f=0.48$; A) $\chi_s = 0$ and B) $\chi_s = 0.5$. $x=1000$, $z=3$ for all curves.} \label{Fig_Conf_entr2}
\end{figure}

Stiffness of polymers strongly affects the behavior of configurational entropy near the surface. As expected, the bulk density of stiffer polymers is reduced, which results in smaller number of configurational states. Figure \ref{Fig_Conf_entr2}A demonstrates that purely repulsive surface produces strictly states with $S_{conf}>0$. However, attractive surface strongly alters the configurational entropy profile, which is demonstrated in Fig. \ref{Fig_Conf_entr2}B where one observes appearance of negative entropies near surface with increase of polymer stiffness.

\begin{figure}
\epsfxsize=3.2in \epsffile{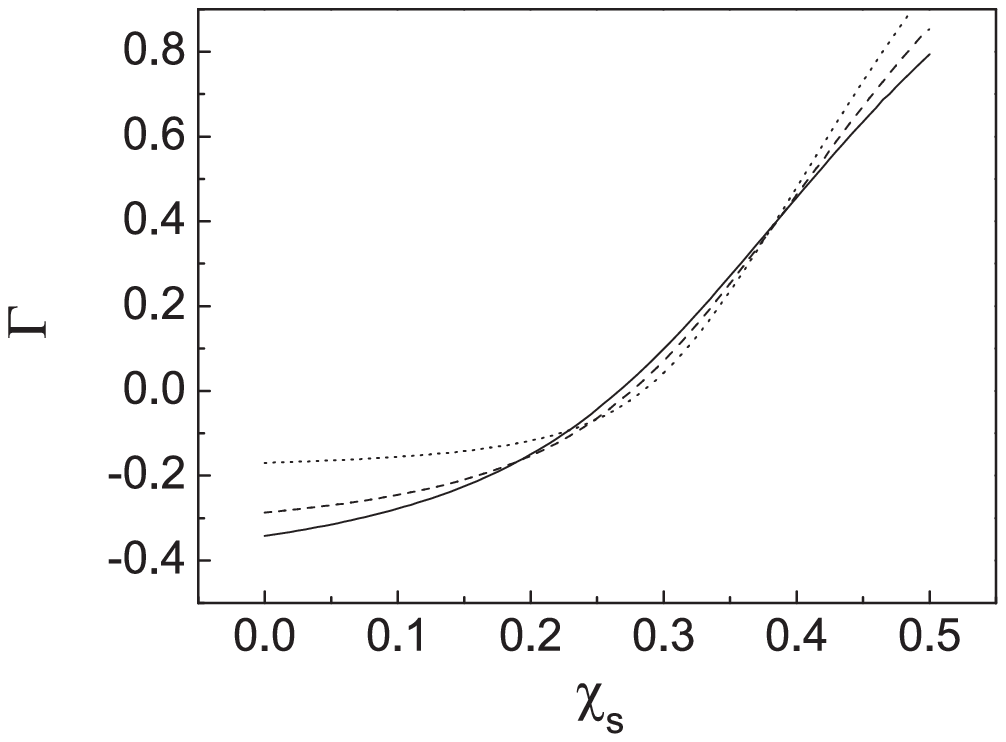}%
\caption{Surface depletion as a function of adsorption energy for 3 values of gauche state probability: solid line - $f=0.5$, dash line - $f=0.49$, dotted line - $f=0.48$. $x=1000$, $z=3$ is the same for all curves.} \label{Fig_Gamma}
\end{figure}

Dependence of excess density $\Gamma$ is demonstrated in Fig. \ref{Fig_Gamma}. According to the experimental \cite{DecreaseTg5} and theoretical \cite{Stepanow} results, higher stiffness macromolecules desorb from a surface less than the flexible polymers, explained by more globular shape of flexible macromolecules, which results in decrease of number of segment-surface connections and, hence, in reduced surface depletion, which is observed in Fig. \ref{Fig_Gamma} for weak adsorption, $\chi_s < 0.25$. On the other hand, MC simulations \cite{Sumithra} of adsorption of a single semiflexible polymer chain onto a planar homogeneous surface showed that stiffer chains adsorb more onto the surface, which creates larger depletion in terms of $\Gamma$ than for flexible chains. The latter agrees with our results for strong adsorption, $\chi_s > 0.4$. Specific region $\Gamma \approx 0$ indicates weak dependence of density profile on polymer semiflexibility. The explanation of this effect could be that, when the entropic density depletion near the surface is nearly removed by the attractive interaction, chains feel no surface for $\chi_s \simeq 0.25 .. 0.4$, which should lead to much weaker contribution of semiflexibility to $\Gamma$ as found.

\section{Conclusion} \label{SecConcl}
The model makes use of the Bethe-Peierls approximation and utilizes the concept of fraction of bonds in the melt in the gauche state derived by Flory. In order to find the local density of melt near surface, the conformational properties of single chain are compared with configurations of small clusters. Based on this comparison the expression for chemical potential is obtained. Using the condition of equality between chemical potentials in neighboring layers a set of equations are derived that are analog of the SF theory for adsorption of interacting chains. The resulting model is solved numerically using a simple iterative algorithm. Computation results of the configurational entropy are in qualitative agreement with the experimental observations about possible existence of a glassy layer near hard surfaces in polymer systems. In addition, it makes possible to estimate the role of the energetic and entopic contributions to the free energy of the interface formation in the melt state. Such information is extremely useful in prediction of the critical stress required to disrupt a polymer film adsorbed on a surface. Another possible application is in the area of nanocomposites, where the effective radius of a nanoparticle immersed in a polymer matrix may significantly be  dependent on attractive interaction between polymer matrix and the particle surface, which is important to predict the stability of the dispersions. All these problems we plan to investigate in the forthcoming publications.

\end{document}